\documentstyle[11pt]{article}
\begin{document}
\begin{center}
{\bf Covariant symplectic structure of\\
the complex Monge-Amp\`ere equation}\\[5mm]
{\bf Y. Nutku} \\[5mm]
Feza G\"ursey Institute, P.O. Box 6 \c{C}engelk\"oy 81220 Istanbul,
Turkey\\[5mm]
\end{center}

\noindent {\bf abstract :}

   The complex Monge-Amp\`ere equation is invariant under arbitrary
holomorphic changes of the independent variables with unit
Jacobian. We present its variational formulation where the action
remains invariant under this infinite group. The new Lagrangian
enables us to obtain the first symplectic $2$-form for the complex
Monge-Amp\`ere equation in the framework of the covariant
Witten-Zuckerman approach to symplectic structure. We base our
considerations on a reformulation of the Witten-Zuckerman theory
in terms of holomorphic differential forms. The first closed and
conserved Witten-Zuckerman symplectic $2$-form for the complex
Monge-Amp\`ere equation is obtained in arbitrary dimension and for
all cases elliptic, hyperbolic and homogeneous.

   The connection of the complex Monge-Amp\`ere equation with Ricci-flat
K\"ahler geometry suggests the use of the Hilbert action principle
as an alternative variational formulation. However, we point out
that Hilbert's Lagrangian is a divergence for K\"ahler metrics and
serves as a topological invariant rather than yielding the
Euclideanized Einstein field equations. Nevertheless, since the
Witten-Zuckerman theory employs only the boundary terms in the
first variation of the action, Hilbert's Lagrangian can be used to
obtain the second Witten-Zuckerman symplectic $2$-form. This
symplectic $2$-form vanishes on shell, thus defining a Lagrangian
submanifold. In its derivation the connection of the second
symplectic $2$-form with the complex Monge-Amp\`ere equation is
indirect but we show that it satisfies all the properties required
of a symplectic $2$-form for the complex elliptic, or hyperbolic
Monge-Amp\`ere equation when the dimension of the complex manifold
is $3$ or higher.

   The complex Monge-Amp\`ere equation admits covariant bisymplectic
structure for complex dimension $3$, or higher. However, in the
physically interesting case of $n=2$ we have only one symplectic
$2$-form.

   The extension of these results to the case of complex
Monge-Amp\`ere-Liouville equation is also presented.

\section{Introduction}

   The complex Monge-Amp\`ere equation plays a central role in
physics and mathematics. On a complex manifold $\cal{M}$ of
dimension $n$ it is given by
\begin{equation}
\frac{1}{n!}( \partial \bar{\partial} u )^n = \kappa \, ^* 1,
                  \;\;\;\;\;\;\;\; \kappa = \pm 1, 0
\label{cma}
\end{equation}
where $\partial$ is the holomorphic exterior derivative,
$\bar{\partial}$ is its anti-holomorphic
counter-part, the wedge product
is understood and $^* 1$ denotes the volume element.
It will be referred to as $CMA_n$. Perhaps the more familiar
form of this equation employs the Monge-Amp\`ere determinant
\begin{equation}
\mu \equiv \det u_{i\bar{k}}
=  \frac{1}{n!}\, ^* ( \partial \bar{\partial} u )^n
  = \kappa
\label{madet}
\end{equation}
where
\begin{equation}
u_{i\bar{k}}  \equiv  \frac{\partial^2  u}
{\partial \zeta^i \partial  \bar{\zeta}^k},
\label{matrix2}
\end{equation}
and $^*$ is the Hodge star duality operator. Depending on $\kappa
= \pm 1$ $CMA_n$ is elliptic, or hyperbolic respectively and it is
called homogeneous for $\kappa=0$. Chern, Levine and Nirenberg
\cite{cln} have pointed out that the homogeneous $CMA_n$ is the
fundamental equation in the theory of functions of many complex
variables. The Laplace equation itself is $CMA_1$. In differential
geometry elliptic $CMA_n$ is the equation governing the K\"ahler
potential for metrics with Euclidean signature and vanishing first
Chern class \cite{cc}. This was noted by Calabi \cite{calabi} and
Yau \cite{yau} has given an existence proof. The Riemann curvature
$(1,1)$-form is then self-dual which by the first Bianchi identity
implies Ricci-flatness. These metrics are hyper-K\"ahler
\cite{an}. Furthermore, since Ricci-flat metrics satisfy the
vacuum Einstein field equation with Euclidean signature, $CMA_2$
is of vital interest for general relativistic instantons. The
construction of the Riemannian metric for the gravitational instanton $K3$,
also familiar as Kummer's surface \cite{kummer}, still remains as
the outstanding unsolved problem of elliptic $CMA_2$.

  Recently we have found \cite{n1} that real Monge-Amp\`ere
equations admit rich multi-symplectic structure. In this letter we
shall present the symplectic structure of $CMA_n$ which, in spite
of its radical differences with the real case, also admits
interesting multi-symplectic structure. $CMA_n$ is invariant under
arbitrary holomorphic changes of the independent variables. In the
elliptic and hyperbolic cases we must add the proviso that the
Jacobian must be unity \cite{bw}. Clearly, the symplectic
structure of $CMA_n$ must be covariant under this infinite group.
The usual approach to symplectic structure starts with a choice of
time parameter, thus breaking covariance from the outset. It is
therefore not suitable for a discussion of $CMA_n$. Happily,
Witten \cite{w} and Zuckerman \cite{z} have shown that there
exists a covariant, closed and conserved $2$-form vector density,
the time component of which gives the familiar symplectic
$2$-form. We shall use the Witten-Zuckerman approach to symplectic
structure, reformulating it in order to express everything in
terms of holomorphic and anti-holomorphic differential forms.

Finally, we note that the usual discussion of the symplectic structure
of $CMA_n$, {\it cf} \cite{semmes}, is based on the K\"ahler $(1,1)$-form
\begin{equation}
\omega_K = \frac{1}{2i}  \partial {\bar {\partial} } u
\label{kahler}
\end{equation}
but this is not the relevant object that emerges from our
examination of symplectic structure. Our approach to symplectic
structure will be in the framework of dynamical systems with
infinitely many degrees of freedom \cite{cm}. We shall start with
action principles underlying $CMA_n$ and from the Lagrangian
derive the corresponding Witten-Zuckerman symplectic $2$-form. The
results we shall present for $CMA_n$ are quite different from the
usual considerations using the K\"ahler $(1,1)$-form.

\section{First symplectic structure of $CMA_n$}

   Everything that is of interest is derivable from a
variational principle \cite{voltaire} $ \delta I = 0$,
\begin{equation}
I = \int L   \label{action}
\end{equation}
where $L$ is the Lagrangian volume form. For $CMA_n$ the Lagrangian is an
$(n,n)$-form
\begin{equation}
 L  =  \frac{1}{(n+1)!} \, \partial u \,  \bar{\partial} u \,
( \partial \bar{\partial} u )^{n-1} + \kappa \, u \, ^* 1
\label{lagrange1}
\end{equation}
and it can be verified directly that first variation of the action
(\ref{action}) yields $CMA_n$. This Lagrangian has not been
considered before and the easiest way of remembering it is through
its determinantal structure
\begin{equation}
^* L =  \det \left| \begin{array}{cc}
  0 & u_{i} \\ u_{\bar{k}} & u_{i\bar{k}}
\end{array} \right|  + \kappa \, u
\label{lagdet}
\end{equation}
which should be compared to the Monge-Amp\`ere determinant
(\ref{madet}). The Lagrangian (\ref{lagrange1}) is also a linear
combination of the zeroth and second order differential invariants
in the prolongation structure and group foliation of $CMA_n$
\cite{ns}.

  We shall use a reformulation of the Witten-Zuckerman theory in terms of
holomorphic and anti-holomorphic differential forms to obtain the symplectic
structure of $CMA_n$. The first variation of the Lagrangian (\ref{lagrange1})
yields
\begin{equation}
\delta \, L = \; \partial \alpha \; + (-1)^n \bar{\partial}  \bar{\alpha}
+ (  \kappa - \mu  ) \, \delta u  \; ^* 1
\label{vary1}
\end{equation}
where $\alpha$ is given by
\begin{equation}
\alpha =  \frac{1}{2 n!} \; \delta u \;  \bar{\partial} u \;
      ( \partial  \bar{\partial} u )^{n-1}
+  \frac{n-1}{2 (n+1)!} \; \bar{\partial} \delta u \;
          \partial u \;  \bar{\partial} u \;
           ( \partial  \bar{\partial} u )^{n-2}
\label{alpha1}
\end{equation}
which is an $(n-1,n)$-form on $\Lambda^{n-1,n} (\cal{M})$ as well
as a $1$-form on $\Lambda^1 (\cal{F(M)})$, the space of functions
over $\cal{M}$. We note that in Witten-Zuckerman theory of
symplectic structure we require that the Jacobi equation
\begin{equation}
( \partial  \bar{\partial} u )^{n-1} \partial  \bar{\partial} \, \delta u = 0
\label{delta1}
\end{equation}
must be satisfied in addition to $CMA_n$ itself.
Then the Witten-Zuckerman symplectic $2$-form is obtained by
applying the exterior-functional derivative to $\alpha$.
We find that the symplectic $2$-form $\omega=\delta \alpha$ is given by
\begin{eqnarray}
\omega & = &  \frac{1}{2 n!} \;
             \delta u \; (\partial  \bar{\partial} u)^{n-1}
              \;  \bar{\partial} \delta u \nonumber \\
& &+   \frac{n-1}{2 n!} \; \delta u \;
       \bar{\partial} u \;(\partial  \bar{\partial} u)^{n-2}
\partial  \bar{\partial} \delta u \nonumber \\
& &- \frac{n-1}{2 (n+1)!} \; \bar{\partial} \delta u \;
            \bar{\partial} u  \; (\partial  \bar{\partial} u)^{n-2}
                     \; \partial \delta u  \label{omega1}  \\
&& + \frac{n-1}{2 (n+1)!} \; \bar{\partial} \delta u \;
  \partial u \;  (\partial  \bar{\partial} u)^{n-2}
        \;  \bar{\partial} \delta u  \nonumber \\
&&+ \frac{(n-1) (n-2)}{2 (n+1)!} \;  \bar{\partial} \delta u \; \partial u \;
  \bar{\partial} u \; (\partial  \bar{\partial} u)^{n-3} \;
   \partial  \bar{\partial} \delta u . \nonumber
\end{eqnarray}
It can be directly verified that $\omega$ satisfies the closed
\begin{equation}
  \delta \, \omega = 0
\label{closed}
\end{equation}
and conserved
\begin{equation}
 \partial \omega + (-1)^n \bar{\partial} \bar{\omega} = 0
\label{cons}
\end{equation}
properties of the Witten-Zuckerman symplectic structure on
$\Lambda^2 (\cal{F(M)})$ $\otimes$
$ \left[ \right.$ $\Lambda^{n-1,n} (\cal{M})$ $\oplus$
$\Lambda^{n,n-1} (\cal{M})$ $\left. \right]$.
The symmetry group of $CMA_n$ is the group of holomorphic changes of
the independent variables with unit Jacobian \cite{bw}. As in the case
of its nearest relative, namely the group of diffeomorphisms,
this is an infinite group. The symplectic $2$-form (\ref{omega1})
should be expressible as a Lie-Poisson structure associated with
this group. That is, it must come from the co-adjoint action
of vector fields belonging to the Lie algebra of the group of
holomorphic changes of the independent variables with unit Jacobian.

\section{Hilbert's variational principle}

   K\"ahler metrics with unit determinant, a requirement identical
to $CMA_n$, are Ricci-flat. Since this condition is the same as
the Euclideanized Einstein field equations for vacuum we are led
to a second variational principle for $CMA_n$  which is simply
Hilbert's Lagrangian \cite{hilbert}. First we recall that the
K\"ahler metric is given by
\begin{equation}
g_{i\bar{k}}  = u_{i\bar{k}}
\label{metric}
\end{equation}
through the definition (\ref{matrix2}) and we note that
\begin{equation}
\mu = \sqrt{g}
\label{rootg}
\end{equation}
must be nonzero. Therefore we must exclude the homogeneous complex
Monge-Amp\`ere equation from this part of the discussion. Then
it is a standard result in differential geometry \cite{lichnerowicz}
that the Ricci tensor for K\"ahler metrics is given by
\begin{equation}
R_{i\bar{k}}  =  (\ln \mu)_{i \bar{k}}
\label{ricci}
\end{equation}
which makes manifest the important role $CMA_n$ plays in K\"ahler
geometry.

The Hilbert Lagrangian density is
\begin{equation}
 {\cal L}_{H}  =  \sqrt{g} \, R
\label{hp}
\end{equation}
where $R=g^{i\bar{k}} R_{i\bar{k}}$ is the scalar of curvature formed
out of the Riemann tensor. From eq.(\ref{ricci}) and the definition
of the contravariant metric,
it can be verified that for K\"ahler metrics the Lagrangian (\ref{hp})
can be written as the $(n,n)$-form
\begin{equation}
 L_H  =  \frac{1}{n!} \,  ( \partial \bar{\partial} u )^{n-1} \,
                \partial \bar{\partial} \ln \mu
\label{lagrange2}
\end{equation}
but this is a divergence. Another way of seeing this divergence
property, which has not been generally remarked on, is through
eq.(\ref{ricci}) and the following remarkable identity:

\noindent
{\it Lemma}

For K\"ahler metrics
\begin{equation}
(\sqrt{g} g^{i\bar{k}})_{\bar{k}}=0 \label{wow}
\end{equation}
is an identity. Proof is by direct calculation which is immediate
through the observation that $ \sqrt{g} g^{i\bar{k}} $ is given by
the cofactors of $u_{i\bar{k}}$.

   Hilbert's Lagrangian $(n,n)$-form (\ref{lagrange2}) can be written as
\begin{equation}
   L_H = \partial \bar{\partial}     \,  Z
\label{ddbhilbert}
\end{equation}
in several different ways
\begin{equation}
   Z= \frac{1}{n!}  \left\{ \begin{array}{l}
            u \,    ( \partial \bar{\partial} u )^{n-2} \,
               \partial \bar{\partial} \ln \mu,    \\
\ln \mu \,   ( \partial \bar{\partial} u )^{n-1},  \\
 \bar{\partial} u \,
( \partial \bar{\partial} u )^{n-1} \,
                \partial \ln \mu,
\end{array}           \right.
\label{div}
\end{equation}
which are alternative statements of the fact that for K\"ahler
metrics the Hilbert Lagrangian is a topological invariant. Its
first variation vanishes identically and we are left with only the
boundary terms. Even though $CMA_n$ does not emerge as the
Euler-Lagrange equations from (\ref{lagrange2}) an examination of
the divergence terms is interesting because it is precisely these
boundary terms in the first variation of the action that are
important in the Witten-Zuckerman construction of the symplectic
$2$-form. For this purpose we write $L_H$ in the form
\begin{eqnarray}
   L_H & = & \partial X + (-1)^n \bar{\partial} \bar{X} \label{xa} \\
  X &=& \frac{1}{2 n!} \,  \bar{\partial} u \,
             ( \partial \bar{\partial} u )^{n-2} \,
                \partial \bar{\partial} \ln \mu  ,           \label{x2}
\end{eqnarray}
skipping other alternatives manifest in eq.(\ref{div}) because
they will yield degenerate results for symplectic structure. From
the boundary terms using $\delta \mu = 0$ by eqs.(\ref{delta1})
and (\ref{rootg}), we find that
\begin{equation}
\omega_{2} =  \bar{\partial} \delta u  \;
   \partial \bar{\partial} \delta u \;
  ( \partial \bar{\partial} u )^{n-3} \, \partial \bar{\partial} \ln \mu,
\;\;\;\;\;\;\;\;\;    n > 2
\label{omega2}
\end{equation}
is the second Witten-Zuckerman symplectic $2$-form on $\Lambda^2
(\cal{F(M)})$ $\otimes$ $ \left[ \right.$ $\Lambda^{n-1,n}$ $
(\cal{M})$ $\oplus$ $\Lambda^{n,n-1} (\cal{M})$ $\left. \right]$.
Physically the most interesting case of complex Monge-Amp\`ere
equation is the case $n=2$ but then $\omega_{2}$ vanishes
identically.

   We have arrived at the symplectic $2$-form (\ref{omega2}) in an
unconventional way. However, the ultimate justification for any
result is a direct check of its properties. Eq.(\ref{omega2})
satisfies all the properties required of a symplectic $2$-form for
$CMA_{n>2}$ excluding the homogeneous case. Namely the check of
the closed and conserved property of $\omega_2$ given by
eqs.(\ref{closed}), (\ref{cons}) is immediate. It is remarkable
that for $n=2$, the physically interesting case, $\omega_2$
vanishes identically. For $n>2$ the symplectic $2$-form vanishes
only on shell. This property defines a Lagrangian submanifold
\cite{am}, \cite{hitchin}. The symplectic $2$-form expressed in
terms of action angle variables vanishes on shell for integrable
dynamical systems and the phase space is reduced to a Lagrangian
submanifold. In eq.(\ref{omega2}) we have the infinite dimensional
analogue of this situation.

\section{Complex Monge-Amp\`ere-Liouville equation}
\label{sec-liouville}

   The results we have presented above can be immediately extended
to the complex Monge-Amp\`ere-Liouville equation, $ \mu = \kappa
\, e^{\Lambda u}$, which will henceforth be referred to as
$CMAL_n$
\begin{equation}
\frac{1}{n!}( \partial \bar{\partial} u )^n = \kappa \, e^{\Lambda
u}\,
 ^* 1
\label{cmal}
\end{equation}
that governs Einstein-K\"ahler metrics satisfying
$R_{i\bar{k}}=\Lambda g_{i\bar{k}}$. The Lagrangian
(\ref{lagrange1}) is now modified to the form
\begin{equation}
 L  =  \frac{1}{(n+1)!} \, \partial u \,  \bar{\partial} u \,
( \partial \bar{\partial} u )^{n-1} +
    \frac{\kappa}{\Lambda} \left( e^{\Lambda u} - 1 \right) \, ^* 1
\label{lagrangecmal}
\end{equation}
and it can be verified that $\omega$ in eq.(\ref{omega1}) remains
unchanged as the first Witten-Zuckerman symplectic $2$-form for $CMAL_n$.

   The analysis of the second Witten-Zuckerman symplectic $2$-form
for $CMAL_n$ starts with the appropriate modification of the Hilbert
Lagrangian
\begin{equation}
L_{\Lambda} =   \frac{1}{n!} \left[  ( \partial \bar{\partial} u
)^{n-1} \,
                \partial \bar{\partial} \ln \mu
 - \Lambda \,( \partial \bar{\partial} u )^{n} \right]
\label{lagrange3}
\end{equation}
and once again we find that it is a total divergence.
As in the case of eq.(\ref{div}) it can be written as a
divergence in as many different ways.
The result for the second Witten-Zuckerman symplectic $2$-form is given by
\begin{equation}
\omega_{2 \Lambda} = \bar{\partial} \delta u  \;
   \partial \bar{\partial} \delta u \;
  ( \partial \bar{\partial} u )^{n-3} \, \partial \bar{\partial} \ln \mu
- \Lambda \, \bar{\partial} \delta u  \;
   \partial \bar{\partial} \delta u \;
  ( \partial \bar{\partial} u )^{n-2}
\label{omega2l}
\end{equation}
again with the proviso $n>2$. Just as in the case of $\omega_2$ we
find that $\omega_{2\Lambda}$ also vanishes when the
Einstein-K\"ahler condition is satisfied. This is manifest when we
put $ \mu = \kappa \, e^{\Lambda u}$ in eq.(\ref{omega2l}). Hence,
as in the case of $CMA_n$, the second Witten-Zuckerman symplectic
$2$-form (\ref{omega2l}) for $CMAL_n$ also defines a Lagrangian
submanifold.

\section{Conclusion}

    We have considered the covariant symplectic structure of complex
Monge-Amp\`ere and Monge-Amp\`ere-Liouville equations in the covariant
framework of the Witten-Zuckerman formalism adapted to holomorphic
and anti-holo\-mor\-phic differential forms. The Lagrangians
(\ref{lagrange1}) and (\ref{lagrangecmal}) directly lead to the
non-degenerate Witten-Zuckerman symplectic $2$-form (\ref{omega1})
in arbitrary dimension and for all cases elliptic, hyperbolic and
homogeneous.

   In order to prove the complete integrability of $CMA_n$ we need
two such symplectic $2$-forms and use the theorem of Magri
\cite{magri}. To this end we considered the Hilbert action
principle (\ref{lagrange2}) and (\ref{lagrange3}) for
Euclideanized vacuum Einstein field equations which are satisfied
by virtue of $CMA(L)_n$ and arrived at the symplectic $2$-forms
(\ref{omega2}) and (\ref{omega2l}). Since Hilbert's Lagrangian is
a divergence for K\"ahler metrics, it serves as a topological
invariant rather than a Lagrangian for the Euclideanized Einstein
field equations. Nevertheless, we were able to obtain the
symplectic $2$-forms (\ref{omega2}) and (\ref{omega2l}) because
only the boundary terms in the first variation of the action play
a significant role in the Witten-Zuckerman construction. For $ n>
2 $ they satisfy all the properties required of a symplectic
$2$-form for the complex elliptic, or hyperbolic
Monge-Amp\`ere-(Liouville) equation: $CMA(L)_{n>2}$ admits
bisymplectic structure.

\section{Acknowledgement}

I thank H. G\"umral, A. S. Fokas and M. B. Sheftel' for interesting
conversations. I thank also the referees of this paper for remarks which
made me clarify the meaning of some points and for raising the question
about Lie-Poisson structure.

\end{document}